\pdfoutput=1
\documentclass[12pt,a4paper]{article}

\usepackage{ifthen} 
\newboolean{pdflatex}
\setboolean{pdflatex}{true} 

\newboolean{articletitles}
\setboolean{articletitles}{true} 

\newboolean{uprightparticles}
\setboolean{uprightparticles}{false} 

\def\paperauthors{LHCb collaboration} 
\def\paperasciititle{First observation of the B+->Ds+Ds-K+ decay} 
\def\papertitle{First observation of \\the \BuDsDsK decay} 
\def\paperkeywords{{High Energy Physics}, {LHCb}} 
\def\papercopyright{\the\year\ CERN for the benefit of the LHCb collaboration} 
\def\paperlicence{CC BY 4.0 licence}
\def\paperlicenceurl{https://creativecommons.org/licenses/by/4.0/}

\usepackage[top=1in, bottom=1.25in, left=1in, right=1in]{geometry}

\usepackage{microtype}
\usepackage{lineno}  
\usepackage{xspace} 
\usepackage{caption} 

\usepackage{graphicx}  
\usepackage{color}
\usepackage{colortbl}
\graphicspath{{./figs/},{./eps/}} 

\usepackage{amsmath} 
\usepackage{amssymb}
\usepackage{amsfonts}
\usepackage{upgreek} 

\newcommand*\patchAmsMathEnvironmentForLineno[1]{%
\expandafter\let\csname old#1\expandafter\endcsname\csname #1\endcsname
\expandafter\let\csname oldend#1\expandafter\endcsname\csname
end#1\endcsname
 \renewenvironment{#1}%
   {\linenomath\csname old#1\endcsname}%
   {\csname oldend#1\endcsname\endlinenomath}%
}
\newcommand*\patchBothAmsMathEnvironmentsForLineno[1]{%
  \patchAmsMathEnvironmentForLineno{#1}%
  \patchAmsMathEnvironmentForLineno{#1*}%
}
\AtBeginDocument{%
\patchBothAmsMathEnvironmentsForLineno{equation}%
\patchBothAmsMathEnvironmentsForLineno{align}%
\patchBothAmsMathEnvironmentsForLineno{flalign}%
\patchBothAmsMathEnvironmentsForLineno{alignat}%
\patchBothAmsMathEnvironmentsForLineno{gather}%
\patchBothAmsMathEnvironmentsForLineno{multline}%
\patchBothAmsMathEnvironmentsForLineno{eqnarray}%
}

\usepackage{hyperxmp}

\usepackage[pdftex,
            pdfauthor={\paperauthors},
            pdftitle={\paperasciititle},
            pdfkeywords={\paperkeywords},
            pdfcopyright={Copyright (C) \papercopyright},
            pdflicenseurl={\paperlicenceurl}]{hyperref}

\usepackage[colorinlistoftodos,textsize=scriptsize]{todonotes}

\usepackage[bottom,flushmargin,hang,multiple]{footmisc}

\usepackage[all]{hypcap} 

\usepackage{xspace} 
\usepackage{upgreek}

\def\lhcb   {\mbox{LHCb}\xspace}

\def\MagUp {\mbox{\em Mag\kern -0.05em Up}\xspace}

\ifthenelse{\boolean{uprightparticles}}%
{

 \def\Ppi         {\ensuremath{\uppi}\xspace}

 \def\Ppsi        {\ensuremath{\uppsi}\xspace}

 \def\PDelta      {\ensuremath{\Delta}\xspace}                 
 \def\PXi         {\ensuremath{\Xi}\xspace}                 
 \def\PLambda     {\ensuremath{\Lambda}\xspace}                 
 \def\PSigma      {\ensuremath{\Sigma}\xspace}                 
 \def\POmega      {\ensuremath{\Omega}\xspace}                 
 \def\PUpsilon    {\ensuremath{\Upsilon}\xspace}
 \let\oldPi\Pi
 \def\PPi         {\ensuremath{\oldPi}\xspace}

 \def\PB      {\ensuremath{\mathrm{B}}\xspace}                 
                  
 \def\PD      {\ensuremath{\mathrm{D}}\xspace}

 \def\PJ      {\ensuremath{\mathrm{J}}\xspace}                 
 \def\PK      {\ensuremath{\mathrm{K}}\xspace}

 \def\Pb      {\ensuremath{\mathrm{b}}\xspace}                 
 \def\Pc      {\ensuremath{\mathrm{c}}\xspace}

 \def\Pi      {\ensuremath{\mathrm{i}}\xspace}

 \def\Ps      {\ensuremath{\mathrm{s}}\xspace}

 \def\thebaroffset{0.0em}
}
{

 \def\Ppi         {\ensuremath{\pi}\xspace}

 \def\Ppsi        {\ensuremath{\psi}\xspace}                 
                  
 \mathchardef\PDelta="7101
 \mathchardef\PXi="7104
 \mathchardef\PLambda="7103
 \mathchardef\PSigma="7106
 \mathchardef\POmega="710A
 \mathchardef\PUpsilon="7107
 \mathchardef\PPi="7105
                  
 \def\PB      {\ensuremath{B}\xspace}                 
                  
 \def\PD      {\ensuremath{D}\xspace}

 \def\PJ      {\ensuremath{J}\xspace}                 
 \def\PK      {\ensuremath{K}\xspace}

 \def\Pb      {\ensuremath{b}\xspace}                 
 \def\Pc      {\ensuremath{c}\xspace}

 \def\Pi      {\ensuremath{i}\xspace}

 \def\Ps      {\ensuremath{s}\xspace}

 \def\thebaroffset{0.18em}
}
\newcommand{\offsetoverline}[2][\thebaroffset]{\kern #1\overline{\kern -#1 #2}}%

\makeatletter
\ifcase \@ptsize \relax
  \newcommand{\miniscule}{\@setfontsize\miniscule{4}{5}}
\or
  \newcommand{\miniscule}{\@setfontsize\miniscule{5}{6}}
\or
  \newcommand{\miniscule}{\@setfontsize\miniscule{5}{6}}
\fi
\makeatother

\DeclareRobustCommand{\optbar}[1]{\shortstack{{\miniscule (\rule[.5ex]{1.25em}{.18mm})}
  \\ [-.7ex] $#1$}}

\def\ellm       {{\ensuremath{\ell^-}}\xspace}
\def\ellp       {{\ensuremath{\ell^+}}\xspace}

\def\squark    {{\ensuremath{\Ps}}\xspace}
\def\squarkbar {{\ensuremath{\overline \squark}}\xspace}

\def\cquark    {{\ensuremath{\Pc}}\xspace}
\def\cquarkbar {{\ensuremath{\overline \cquark}}\xspace}
\def\ccbar     {{\ensuremath{\cquark\cquarkbar}}\xspace}
\def\bquark    {{\ensuremath{\Pb}}\xspace}
\def\bquarkbar {{\ensuremath{\overline \bquark}}\xspace}

\def\pion   {{\ensuremath{\Ppi}}\xspace}
\def\piz    {{\ensuremath{\pion^0}}\xspace}
\def\pip    {{\ensuremath{\pion^+}}\xspace}

\def\pipm   {{\ensuremath{\pion^\pm}}\xspace}

\def\kaon    {{\ensuremath{\PK}}\xspace}

\def\KorKbar {\kern \thebaroffset\optbar{\kern -\thebaroffset \PK}{}\xspace}

\def\Kp      {{\ensuremath{\kaon^+}}\xspace}
\def\Km      {{\ensuremath{\kaon^-}}\xspace}
\def\Kpm     {{\ensuremath{\kaon^\pm}}\xspace}
\def\Kmp     {{\ensuremath{\kaon^\mp}}\xspace}

\def\Dbar    {{\ensuremath{\offsetoverline{\PD}}}\xspace}
\def\D       {{\ensuremath{\PD}}\xspace}

\def\DorDbar {\kern \thebaroffset\optbar{\kern -\thebaroffset \PD}\xspace}

\def\Dp      {{\ensuremath{\D^+}}\xspace}
\def\Dm      {{\ensuremath{\D^-}}\xspace}
\def\Dpm     {{\ensuremath{\D^\pm}}\xspace}

\def\DpDm    {\ensuremath{\Dp {\kern -0.16em \Dm}}\xspace}

\def\Ds      {{\ensuremath{\D^+_\squark}}\xspace}
\def\Dsp     {{\ensuremath{\D^+_\squark}}\xspace}
\def\Dsm     {{\ensuremath{\D^-_\squark}}\xspace}
\def\Dspm    {{\ensuremath{\D^{\pm}_\squark}}\xspace}

\def\B       {{\ensuremath{\PB}}\xspace}

\def\BorBbar {\kern \thebaroffset\optbar{\kern -\thebaroffset \PB}\xspace}

\def\Bd      {{\ensuremath{\B^0}}\xspace}

\def\BdorBdbar {\kern \thebaroffset\optbar{\kern -\thebaroffset \Bd}\xspace}
\def\Bu      {{\ensuremath{\B^+}}\xspace}

\def\Bp      {{\ensuremath{\Bu}}\xspace}

\def\Bs      {{\ensuremath{\B^0_\squark}}\xspace}

\def\BsorBsbar {\kern \thebaroffset\optbar{\kern -\thebaroffset \Bs}\xspace}

\def\jpsi     {{\ensuremath{{\PJ\mskip -3mu/\mskip -2mu\Ppsi}}}\xspace}

\def\Y#1S{\ensuremath{\PUpsilon{(#1S)}}\xspace}

\def\LorLbar     {\kern \thebaroffset\optbar{\kern -\thebaroffset \PLambda}\xspace}

\newcommand{\decay}[2]{\ensuremath{#1\!\to #2}\xspace} 

\def\to                 {\ensuremath{\rightarrow}\xspace}

\def\AT#1     {\ensuremath{A_{\mathrm{T}}^{#1}}\xspace}           

\def\C#1      {\ensuremath{\mathcal{C}_{#1}}\xspace}                       
\def\Cp#1     {\ensuremath{\mathcal{C}_{#1}^{'}}\xspace}                    
\def\Ceff#1   {\ensuremath{\mathcal{C}_{#1}^{\mathrm{(eff)}}}\xspace}        
\def\Cpeff#1  {\ensuremath{\mathcal{C}_{#1}^{'\mathrm{(eff)}}}\xspace}       
\def\Ope#1    {\ensuremath{\mathcal{O}_{#1}}\xspace}                       
\def\Opep#1   {\ensuremath{\mathcal{O}_{#1}^{'}}\xspace}                    

\newcommand{\nospaceunit}[1]{\ensuremath{\text{#1}}}       
\newcommand{\aunit}[1]{\ensuremath{\text{\,#1}}}       

\newcommand{\tev}{\aunit{Te\kern -0.1em V}\xspace}
\newcommand{\gev}{\aunit{Ge\kern -0.1em V}\xspace}
\newcommand{\mev}{\aunit{Me\kern -0.1em V}\xspace}
\newcommand{\kev}{\aunit{ke\kern -0.1em V}\xspace}
\newcommand{\ev}{\aunit{e\kern -0.1em V}\xspace}
 
\newcommand{\mevc}{\ensuremath{\aunit{Me\kern -0.1em V\!/}c}\xspace}
\newcommand{\gevc}{\ensuremath{\aunit{Ge\kern -0.1em V\!/}c}\xspace}
\newcommand{\mevcc}{\ensuremath{\aunit{Me\kern -0.1em V\!/}c^2}\xspace}
\newcommand{\gevcc}{\ensuremath{\aunit{Ge\kern -0.1em V\!/}c^2}\xspace}

\def\mum  {\ensuremath{\,\upmu\nospaceunit{m}}\xspace}

\def\fb   {\ensuremath{\aunit{fb}}\xspace}
\def\invfb   {\ensuremath{\fb^{-1}}\xspace}

\newcommand{\chisq}{\ensuremath{\chi^2}\xspace}

\def\gsim{{~\raise.15em\hbox{$>$}\kern-.85em
          \lower.35em\hbox{$\sim$}~}\xspace}
\def\lsim{{~\raise.15em\hbox{$<$}\kern-.85em
          \lower.35em\hbox{$\sim$}~}\xspace}

\def\sPlot{\mbox{\em sPlot}\xspace}

\def\pt         {\ensuremath{p_{\mathrm{T}}}\xspace}

\def\ptot       {\ensuremath{p}\xspace}

\def\evtgen     {\mbox{\textsc{EvtGen}}\xspace}

\def\geant      {\mbox{\textsc{Geant4}}\xspace}

\def\photos     {\mbox{\textsc{Photos}}\xspace}

\def\pythia     {\mbox{\textsc{Pythia}}\xspace}

\def\tell1  {TELL1\xspace}
\def\ukl1   {UKL1\xspace}

\newcommand{\eg}{\mbox{\itshape e.g.}\xspace}
\newcommand{\ie}{\mbox{\itshape i.e.}\xspace}

\newcommand{\lhcborcid}[1]{\href{https://orcid.org/#1}{\hspace*{0.1em}\raisebox{-0.45ex}{\includegraphics[width=1em]{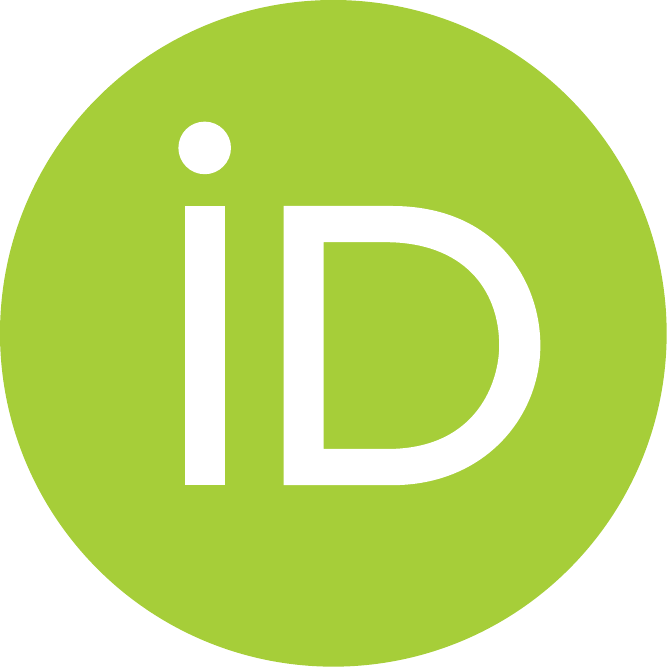}}}}

\def\BuDsDsK       {\ensuremath{\mbox{\decay{\Bp}{\Dsp\Dsm\Kp}}}\xspace}
\def\BuDDK         {\ensuremath{\mbox{\decay{\Bp}{\Dp\Dm\Kp}}}\xspace}
\def\BuDstDstKst   {\ensuremath{\mbox{$B\to D^{(*)}_{(s)}\Dbar^{(*)}_{(s)}K^{(*)}$}}\xspace}

\def\DsKKpi        {\ensuremath{\mbox{\Ds\to\Km\Kp\pip}}\xspace}
\def\DKpipi        {\ensuremath{\mbox{\Dp\to\Km\pip\pip}}\xspace}

\def\msDsDs        {\ensuremath{\mbox{$m^2(\Dsp\Dsm)$}}\xspace}
\def\msDsK        {\ensuremath{\mbox{$m^2(\Dsm\Kp)$}}\xspace}

\def\msDD        {\ensuremath{\mbox{$m^2(\Dp\Dm)$}}\xspace}
\def\msDK        {\ensuremath{\mbox{$m^2(\Dm\Kp)$}}\xspace}

\def\Rbr         {\ensuremath{{\cal R}}\xspace}
\def\Brsig          {\ensuremath{\mbox{${\cal B}$(\BuDsDsK)}}\xspace}
\def\BrDs          {\ensuremath{\mbox{${\cal B}$(\DsKKpi)}}\xspace}
\def\Brcon          {\ensuremath{\mbox{${\cal B}$(\BuDDK)}}\xspace}
\def\BrD          {\ensuremath{\mbox{${\cal B}$(\DKpipi)}}\xspace}

\def\Rfrac         {\ensuremath{{\Brsig\over\Brcon}}\xspace}

\def\Nsigcorr        {\ensuremath{N^{\rm corr}_{\rm sig}}\xspace}
\def\Nconcorr           {\ensuremath{N^{\rm corr}_{\rm norm}}\xspace}

\def\esigi             {\ensuremath{\epsilon_{{\rm sig},\, i}}\xspace}
\def\econi             {\ensuremath{\epsilon_{{\rm norm},\, i}}\xspace}
\def\wsigi             {\ensuremath{w_{{\rm sig},\, i}}\xspace}
\def\wconi             {\ensuremath{w_{{\rm norm},\, i}}\xspace}

\newcommand{\phsys}{\phantom{~(6.2)}}

\usepackage{cite} 
\usepackage{mciteplus}

\begin{document}

\renewcommand{\thefootnote}{\fnsymbol{footnote}}
\setcounter{footnote}{1}



\begin{titlepage}
\pagenumbering{roman}

\vspace*{-1.5cm}
\centerline{\large EUROPEAN ORGANIZATION FOR NUCLEAR RESEARCH (CERN)}
\vspace*{1.5cm}
\noindent
\begin{tabular*}{\linewidth}{lc@{\extracolsep{\fill}}r@{\extracolsep{0pt}}}
\ifthenelse{\boolean{pdflatex}}
{\vspace*{-1.5cm}\mbox{\!\!\!\includegraphics[width=.14\textwidth]{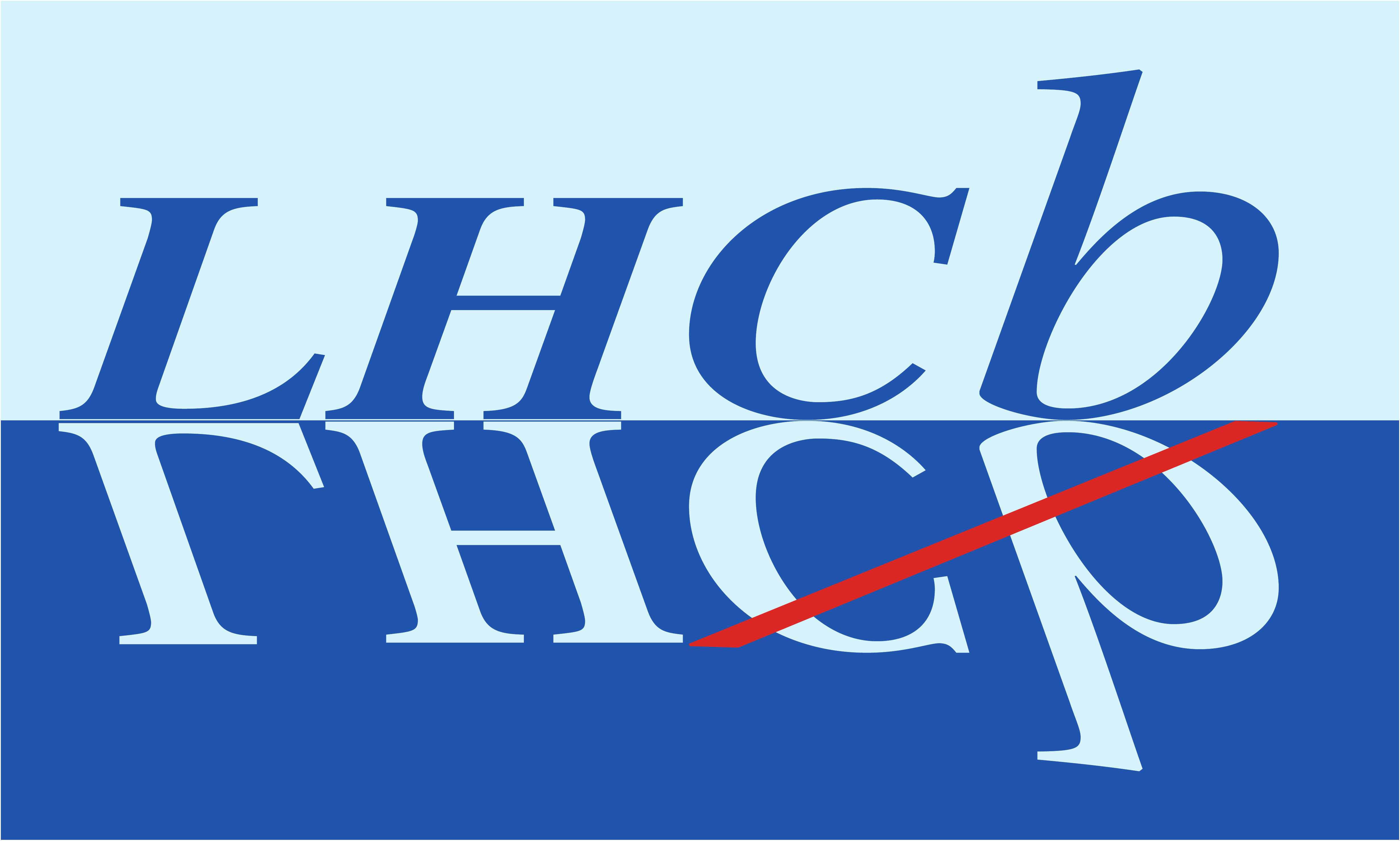}} & &}%
{\vspace*{-1.2cm}\mbox{\!\!\!\includegraphics[width=.12\textwidth]{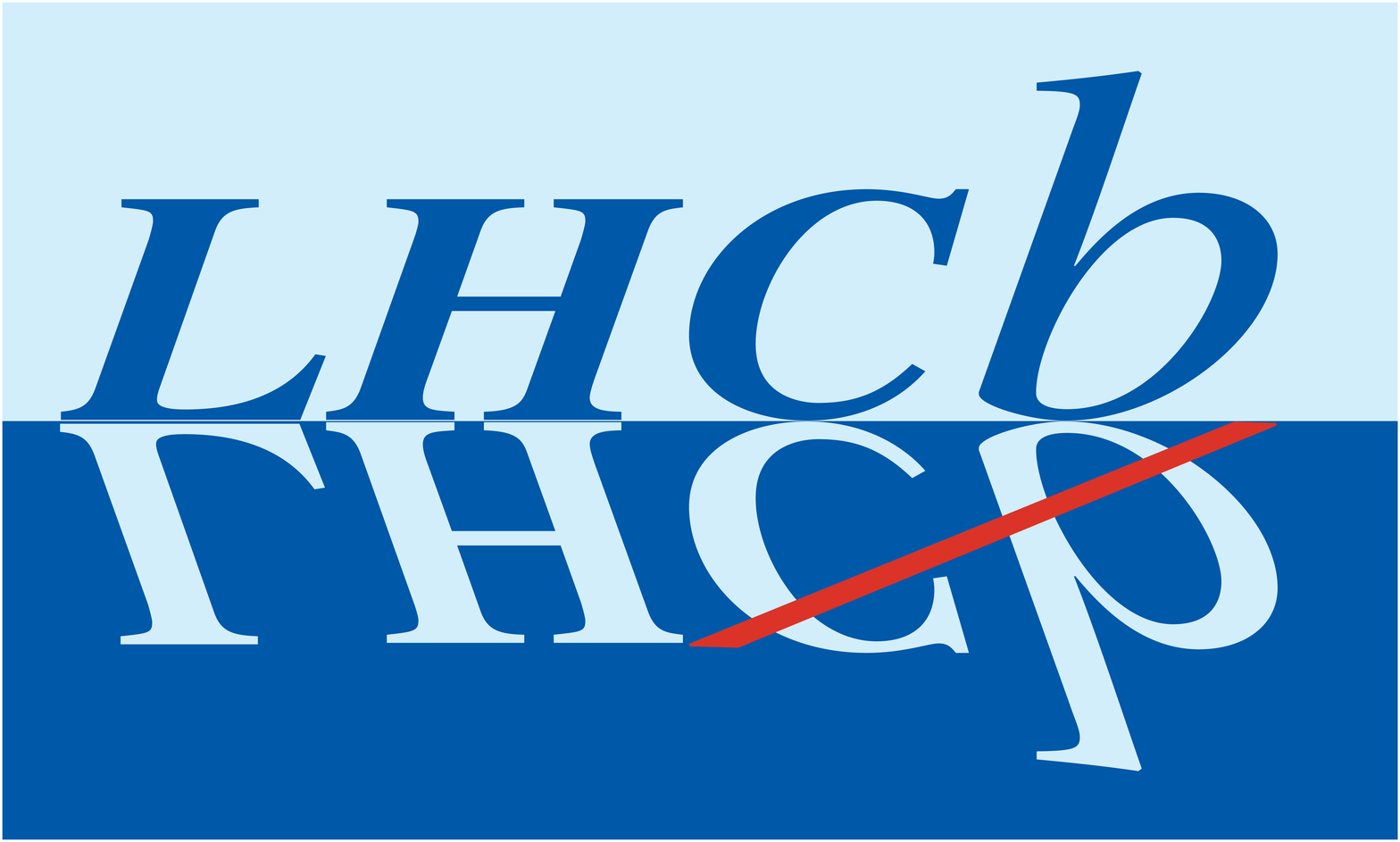}} & &}%
\\
 & & CERN-EP-2022-205 \\  
 & & LHCb-PAPER-2022-019 \\  
 & & 05 Sep 2023 \\ 
 & & \\
\end{tabular*}

\vspace*{4.0cm}

{\normalfont\bfseries\boldmath\huge
\begin{center}
  \papertitle 
\end{center}
}

\vspace*{2.0cm}

\begin{center}
\paperauthors
\footnote{Authors are listed at the end of this paper.}
\end{center}

\vspace{\fill}

\begin{abstract}
  \noindent
  The \BuDsDsK decay is observed for the first time using proton-proton collision data collected by the LHCb detector at centre-of-mass energies of 7, 8 and 13\tev, corresponding to an integrated luminosity of $9\invfb$.
  Its branching fraction  relative to that of the \BuDDK decay is measured to be
  $$\Rfrac = 0.525 \pm 0.033 \pm 0.027 \pm 0.034 ,$$
  where the first uncertainty is statistical, the second systematic, and the third is due to the uncertainties on the branching fractions of the $\Dspm\to\Kmp\Kpm\pipm$ and $\Dpm\to\Kmp\pipm\pipm$ decays.
  This measurement fills an experimental gap in the knowledge of the family of Cabibbo--favoured $\bquarkbar \to \cquarkbar \cquark \squarkbar$ transitions and opens the path for  unique studies of spectroscopy in future.
\end{abstract}

\vspace*{1.0cm}

\begin{center}
    Published in Phys. Rev. \textbf{D108}~(2023) 034012

\end{center}

\vspace{\fill}

{\footnotesize 
\centerline{\copyright~\papercopyright. \href{\paperlicenceurl}{\paperlicence}.}}
\vspace*{2mm}

\end{titlepage}


\newpage
\setcounter{page}{2}
\mbox{~}


\renewcommand{\thefootnote}{\arabic{footnote}}
\setcounter{footnote}{0}

\cleardoublepage


\pagestyle{plain} 
\setcounter{page}{1}
\pagenumbering{arabic}



\section{Introduction}
\label{sec:introduction}

\begin{figure}[bp]
\centering
\includegraphics[width=1.0\textwidth]{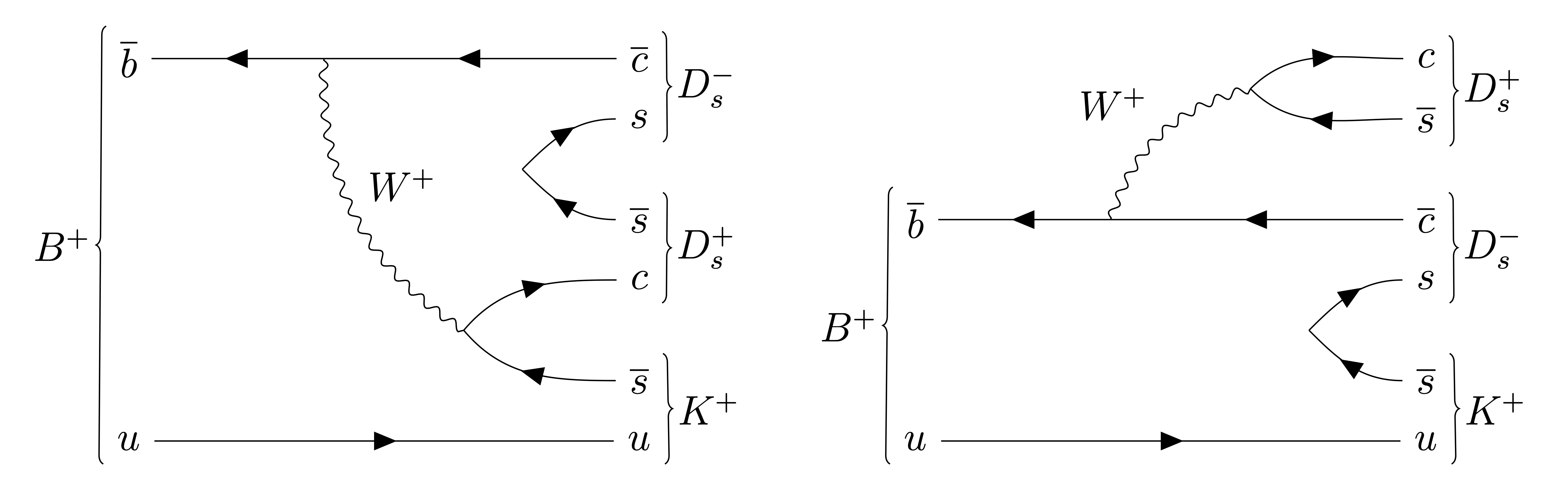}
\caption{Feynman diagrams contributing to the \BuDsDsK decay.}
\label{fig:feynman}
\end{figure}

The family of \BuDstDstKst decays
proceeds at the quark level via the Cabibbo-favoured \mbox{$\bquarkbar \to \cquarkbar \cquark \squarkbar$} transition.\footnote{The inclusion of charge-conjugate modes is always implied and natural units with $\hbar=c=1$ are used throughout the paper.}
Such decays provide an excellent laboratory for investigations of open- and hidden-charm meson spectroscopy~\cite{Aubert:2007rva,Belle:2007hht,Belle:2011ad,Lees:2014abp,LHCb-PAPER-2020-024,LHCb-PAPER-2020-025,LHCb-PAPER-2020-034}, 
covering both conventional and exotic states.
Additionally, measurements of the amplitude structures in the $D_{(s)}^{(*)} \Dbar_{(s)}^{(*)}$ system of these decays can offer important information to the theoretical calculations of the charm-loop contributions to $b\to s \ellp \ellm$ processes that are sensitive to physics beyond the Standard Model~\cite{Khodjamirian:2010vf}. 

There have already been experimental studies of $B\to D^{(*)}\bar{D}^{(*)} K^{(*)} $ decays by the ALEPH, BaBar, Belle and LHCb collaborations~\cite{Barate:1998ch,Aubert:2003jq,Aubert:2007rva,Belle:2007hht,BaBar:2010tqo,Belle:2011ad,Lees:2014abp,LHCb-PAPER-2020-015,LHCb-PAPER-2020-006,LHCb-PAPER-2020-024,LHCb-PAPER-2020-025,LHCb-PAPER-2020-034}, not only making observations of some of these decay channels, but also leading to discoveries of new resonances. 
However, all such measurements to date focus only on $B$ decays with a $D^{(*)} \bar{D}^{(*)}$ pair in the final state. Decays involving the $D_s^{(*)+} D_s^{(*)-}$ pair, \eg\ \BuDsDsK, have never been explored. 
The \Dsp\Dsm system is attractive as it provides a unique insight into the \mbox{charmonium(-like)} spectroscopy.
Conventional charmonium mesons with natural spin~($J$), parity~($P$), and charge-parity~($C$) quantum numbers~($J^{PC}=0^{++},1^{--}, 2^{++}, \cdots$), \eg\ $\psi(4040)$~\cite{PDG2020}, are expected to predominantly decay into the $D^{(*)} \bar{D}^{(*)}$ final state. 
In contrast, charmonium-like hadrons with the \mbox{$\cquark\cquarkbar\squark\squarkbar$} constituents could have a larger partial decay width to the \Dsp\Dsm final state than that to $D^{(*)} \bar{D}^{(*)}$. 
A few candidates for $\ccbar\squark\squarkbar$ states have been observed in the $\jpsi \phi$ final state from \mbox{$\Bp\to\jpsi\phi\Kp$} decays~\cite{LHCb-PAPER-2020-044}. They might also decay into $\Dsp\Dsm$ and contribute to the $\BuDsDsK$ decay.

This paper presents the first observation of the \BuDsDsK decay. The contributing tree-level Feynman diagrams of this decay are shown in Fig.~\ref{fig:feynman}.  
The branching fraction of the \BuDsDsK decay is measured relative to that of the normalisation channel \BuDDK.  The two channels have a similar decay topology, so the corresponding systematic uncertainties on the branching-fraction ratio are expected to largely cancel. 
The measurement uses the proton-proton ($pp$) collision data collected by the LHCb experiment in 2011, 2012 and 2015--2018 at centre-of-mass energies of 7, 8 and 13\tev, respectively, corresponding to a total integrated luminosity of $9\invfb$. 
The branching fraction of the \BuDsDsK decay is an essential input to obtain the partial width information of the near--threshold structure in the \Dsp\Dsm system analysed in the accompanying paper~\cite{LHCb-PAPER-2022-018}.

\section{Detector and simulation}
\label{sec:detector}

The \lhcb detector~\cite{LHCb-DP-2008-001,LHCb-DP-2014-002} is a single-arm forward
spectrometer covering the pseudorapidity range $2<\eta <5$,
designed for the study of particles containing \bquark or \cquark
quarks. The detector includes a high-precision tracking system
consisting of a silicon-strip vertex detector surrounding the $pp$
interaction region~\cite{LHCb-DP-2014-001}, a large-area silicon-strip detector located
upstream of a dipole magnet with a bending power of about
$4{\mathrm{\,Tm}}$, and three stations of silicon-strip detectors and straw
drift tubes~\cite{LHCb-DP-2013-003,LHCb-DP-2017-001} placed downstream of the magnet.
The tracking system provides a measurement of the momentum, \ptot, of charged particles with
a relative uncertainty that varies from 0.5\% at low momentum to 1.0\% at 200\gev.
The minimum distance of a track to a primary $pp$ collision vertex~(PV), the impact parameter, 
is measured with a resolution of $(15+29/\pt)\mum$,
where \pt is the component of the momentum transverse to the beam, in\,\gev.
Different types of charged hadrons are distinguished using information
from two ring-imaging Cherenkov detectors~\cite{LHCb-DP-2012-003}. 
Photons, electrons and hadrons are identified by a calorimeter system consisting of
scintillating-pad and preshower detectors, an electromagnetic
and a hadronic calorimeter. Muons are identified by a
system composed of alternating layers of iron and multiwire
proportional chambers~\cite{LHCb-DP-2012-002}.
    
The online event selection is performed by a trigger~\cite{LHCb-DP-2012-004}, 
which consists of a hardware stage, based on information from the calorimeter and muon systems, followed by a software stage, which applies a full event reconstruction.
At the hardware trigger stage, events are required to have a muon with high \pt or a
hadron, photon or electron with high transverse energy in the calorimeters. 
The hadron can originate from either the studied decay or the rest of the event.
The software trigger requires a two-, three- or four-track
secondary vertex with a significant displacement from any PV. At least one charged particle
must have a transverse momentum $\pt > 1.6\gev$ and be
inconsistent with originating from a PV.
A multivariate algorithm~\cite{BBDT,LHCb-PROC-2015-018} is used for
the identification of secondary vertices consistent with the decay
of a \bquark hadron.

Simulation is required to model the effects of the detector acceptance and the imposed selection requirements for the \BuDsDsK and \BuDDK decays. In the simulation, $pp$ collisions are generated using \pythia~\cite{Sjostrand:2007gs,*Sjostrand:2006za}  with a specific \lhcb configuration~\cite{LHCb-PROC-2010-056}. Decays of unstable particles are described by \evtgen~\cite{Lange:2001uf}, in which final-state radiation is generated using \photos~\cite{davidson2015photos}. The interaction of the generated particles with the detector, and its response, are implemented using the \geant toolkit~\cite{Allison:2006ve, *Agostinelli:2002hh} as described in Ref.~\cite{LHCb-PROC-2011-006}.  The underlying $pp$ interaction is reused multiple times, with an independently generated signal decay for each event~\cite{LHCb-DP-2018-004}.
The particle-identification~(PID) response is not well described in the LHCb simulation, and is corrected to match that in data. The correction is determined from calibration samples using a reweighing approach, the so-called PID transformation~\cite{LHCb-PUB-2016-021,LHCb-DP-2018-001}.

\section{Event selection}
The \BuDsDsK and \BuDDK decays are reconstructed using the  \mbox{$\Dsp\to\Kp\Km\pip$} and \mbox{$\Dp\to\Km\pip\pip$} decay channels, respectively.
The selection starts by choosing well-reconstructed tracks that are 
inconsistent with originating from any PV. 
The selected tracks should have PID information consistent with the corresponding final-state kaons and pions.
Additionally, the opening angle between any two final-state charged tracks must be larger than $0.5\,{\rm mrad}$ to reduce potential reuse of track segments.
The \Dp and \Dsp candidates, obtained by combining \kaon and \pion candidates, are required to have good-quality vertices and their reconstructed masses must lie within $\pm25$\mev of the known masses~\cite{PDG2020}, corresponding to about three times the mass resolution. 
The \Bp candidates are then formed by combining $D_{(s)}^+$, $D_{(s)}^-$ and \Kp candidates.
The \Bp decay vertex must be well reconstructed and significantly displaced from all PVs. The \Bp candidate is required to  have a flight direction pointing back to the PV where it is produced, referred to hereafter as the associated PV.
To reduce the contamination from the non-double-charm~(NDC) \Bp decay candidates that have the same set of the final-state tracks as that of the $\Bp\to D_{(s)}^+ D_{(s)}^- \Kp$ decay but do not involve two real $D_{(s)}^\pm$ mesons, requirements on the distances between the \Bp and $D^{\pm}_{(s)}$ vertices are imposed, taking advantage of the nonzero flight distance of the $D_{(s)}^\pm$ meson. 
A kinematic fit~\cite{Hulsbergen:2005pu} is employed to improve the \Bp mass resolution by constraining the $D^{\pm}_{(s)}$ mass to its known value~\cite{PDG2020} and requiring the \Bp candidate to originate from the associated PV. A second kinematic fit with an additional \Bp mass constraint is utilised when investigating the  distributions of the Dalitz--plot variables in \BuDsDsK and \BuDDK decays.

Gradient Boosted Decision Tree~(BDTG)~\cite{Breiman,AdaBoost,Hocker:2007ht,*TMVA4} classifiers are employed to reduce further background from random combinations of tracks, referred to hereafter as combinatorial background. 
The classifier is trained separately for the \BuDsDsK and \BuDDK channels using the corresponding simulated sample as the signal proxy and candidates in data with a reconstructed \Bp mass of $5360-6000\mev$ as the background proxy. 
The discriminating variables used in the classifier
include PID information of the final-state tracks, kinematic properties and the decay topology of the \Bp and  $D^{\pm}_{(s)}$ candidates.
The selection criterion for the response of each classifier is optimised by maximising the figure of merit, $N_S/\sqrt{N_S+N_B}$, separately for each channel. Here, $N_S$~($N_B$) is the expected signal~(background) yield within $\pm20$\mev of the known \Bp mass~\cite{PDG2020}, estimated as the product of the signal~(background) yield without any BDTG requirement and the signal efficiency~(1$-$background-rejection factor)  for a given criterion. The signal and background yields without the BDTG requirements in each channel are evaluated using a simple fit to the reconstructed \Bp mass distribution with the signal and background probability density functions~(PDF) modelled by a Gaussian function and an exponential function, respectively. The signal efficiency and background-rejection factor are directly obtained from the samples used to test the classifiers.

Multiple candidates, found in a few percent of $pp$ collision events, are mostly due to the duplicated use of tracks from the same event. In each event, the candidate having the smallest \chisq of the kinematic fit without the \Bp mass constraint is retained. 
 
The reconstructed \Dsp\Dsm\Kp and \Dp\Dm\Kp invariant-mass distributions after all the selection requirements are shown in Fig.~\ref{fig:fit_sig}. The mass windows, \mbox{$(5280-80,\, 5280+80)\mev$} and \mbox{$(5280-60,\, 5280+80)\mev$}, for the signal and normalisation channels, respectively, are chosen to exclude the contributions of partially reconstructed background with a missing photon from $D_s^{*\pm}\to D_s^\pm \gamma$ or $D^{*\pm}\to D^\pm \gamma$ decay. Here, the tighter lower-side threshold for the \BuDDK channel than that for the \BuDsDsK channel is found necessary to exclude all visible tails of the partially reconstructed background~\cite{LHCb-PAPER-2020-024,LHCb-PAPER-2020-025}. 
These mass windows are also sufficient to exclude partially reconstructed background with a missing \piz meson from $D_{s}^{*+}\to D_{s}^+ \piz$ or $D^{*+}\to D^+ \piz$ decay, as the reconstructed \Dsp\Dsm\Kp or \Dp\Dm\Kp mass is shifted at least $135\mev$ below the \Bp mass peak~\cite{PDG2020}.
For both the \BuDsDsK signal and \BuDDK normalisation channels, any potential background resulting from the misidentification of a single \kaon or \pion would originate from Cabbibo-suppressed decays and thus is negligible.

\section{Signal extraction}
\label{sec:sig}

\begin{figure}[!tbp]
    \centering
    \includegraphics[width=0.495\textwidth]{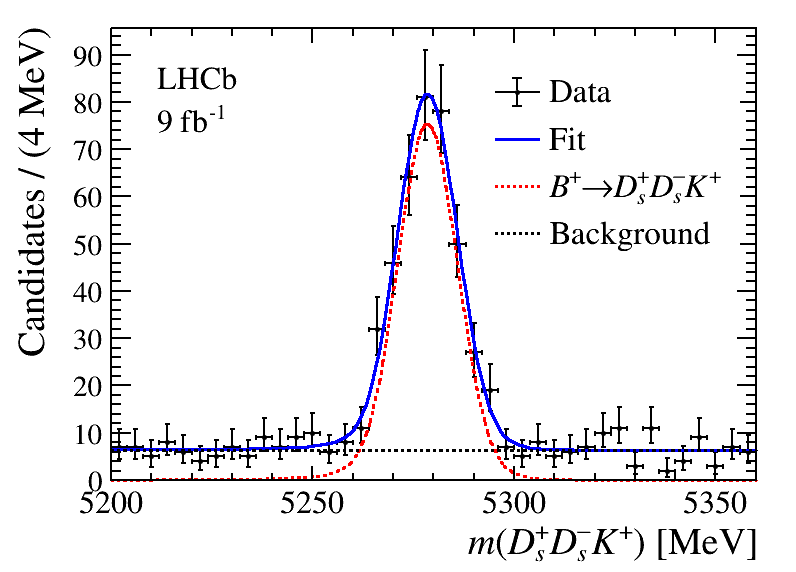}
    \includegraphics[width=0.495\textwidth]{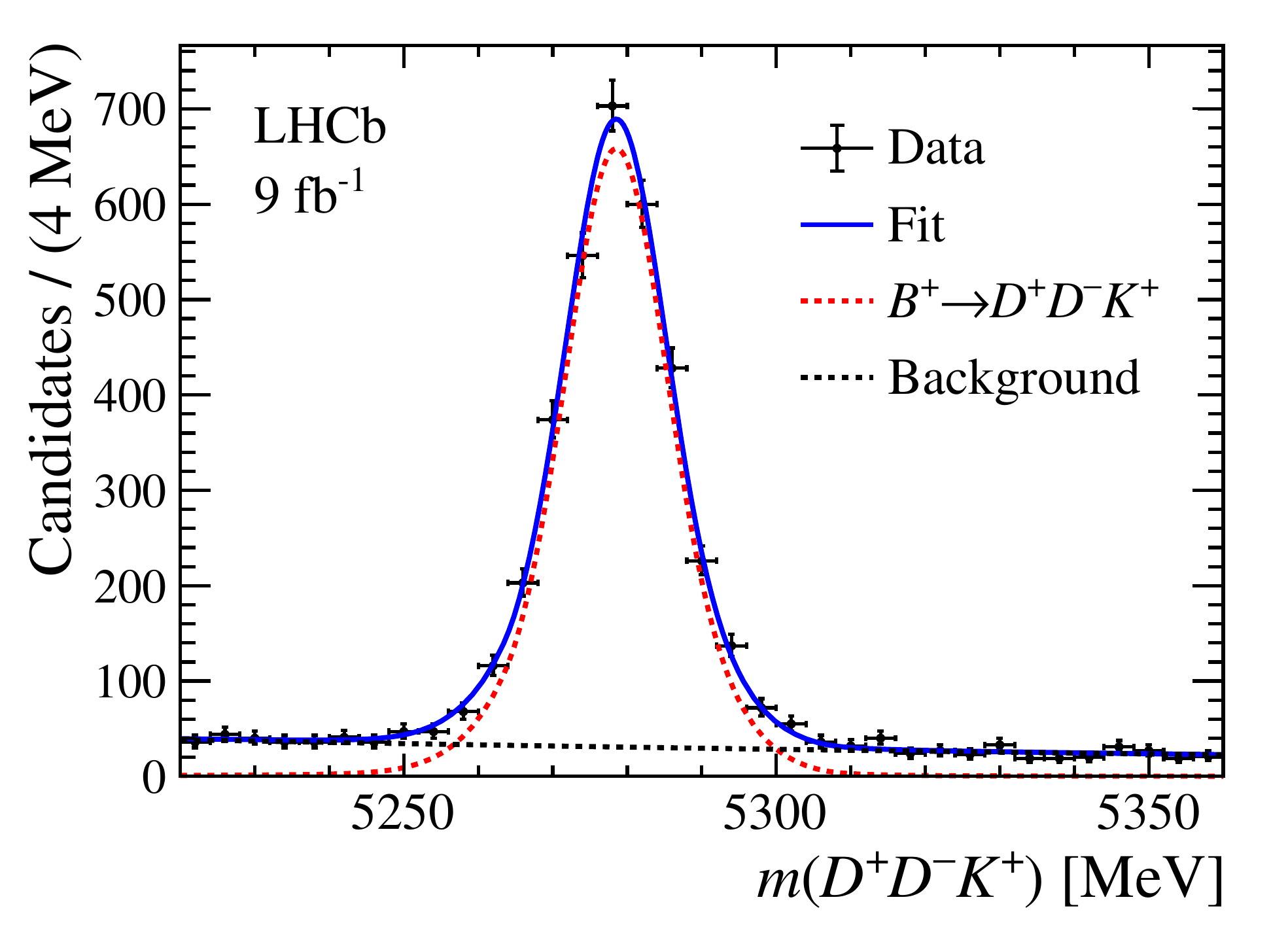}
     \caption{Invariant-mass distributions of reconstructed (left)~\Dsp\Dsm\Kp and (right)~\Dp\Dm\Kp candidates. The fit results are overlaid.}
    \label{fig:fit_sig}
\end{figure}

Unbinned extended maximum-likelihood fits to the mass distributions of the \Bp candidates are performed to determine the signal yields in the \BuDsDsK and \BuDDK channels. 
The signal PDF is a sum of two Crystal Ball~(CB) functions~\cite{Skwarnicki:1986xj} with a common peak position and opposite-side tails. The tail parameters are fixed to the values obtained from the simulated \BuDDK or \BuDsDsK sample and the fractions of the two CB functions are set to 0.5. All other parameters including the peak position, the widths of the two CB functions are free to vary in the fit to data.
The combinatorial background is modelled using an exponential function with its parameter free to vary in the fit. 

The fit results are shown in Fig.~\ref{fig:fit_sig}. 
A signal yield of $N_{\rm sig}=360\pm22$, where the uncertainty is statistical only, is obtained for the \BuDsDsK decay, with a purity in a $\pm20\mev$ window around the peak of approximately $84\%$.  
The statistical significance of the \BuDsDsK decay is evaluated under the assumption that the log-likelihood difference between the fit without the signal component and the default fit follows a \chisq distribution with the number of degrees of freedom being the difference in the number of free parameters of the two fits~\cite{Wilks:1938dza}. 
Given the log-likelihood difference of 646 with four degrees of freedom, the significance is found to be much larger than 10 standard deviations~($\sigma$).
Since the systematic uncertainty discussed in Sec.~\ref{Sec:sys} is not expected to greatly reduce the significance, this result constitutes the first observation of the \BuDsDsK decay.
The signal yield in the \BuDDK channel is obtained as $N_{\rm norm}=3215\pm65$, where the uncertainty is statistical, and the purity is around 91\% using the same mass window definition. 
The \BuDDK signal yield is larger and the corresponding purity is lower compared with those in Refs.~\cite{LHCb-PAPER-2020-024,LHCb-PAPER-2020-025}, reflecting that looser selection criteria are imposed in this paper.

\begin{figure}[!tbp]
    \centering
    \includegraphics[width=0.495\textwidth]{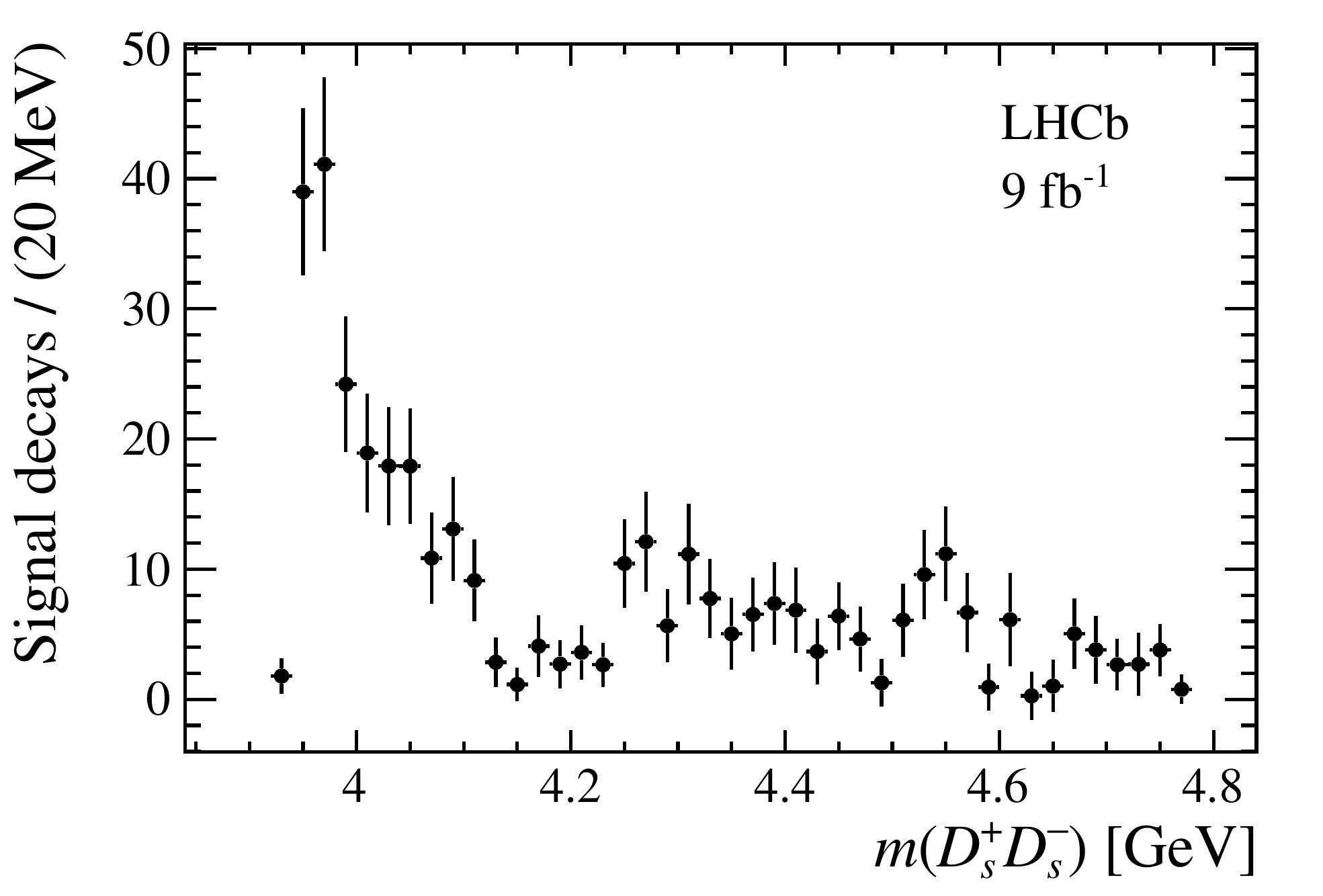}\\
    \includegraphics[width=0.495\textwidth]{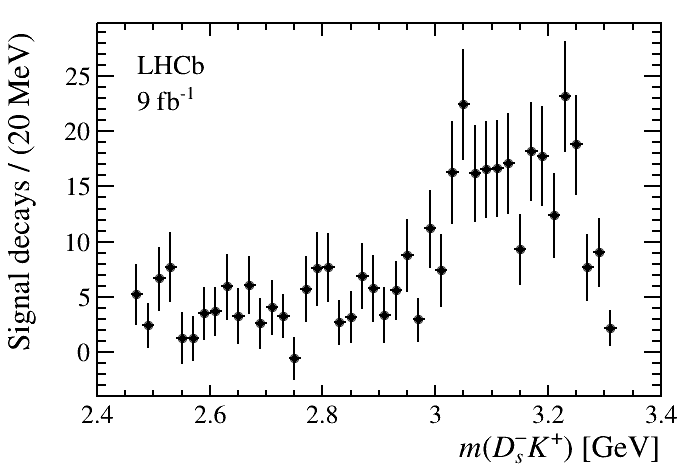}
    \includegraphics[width=0.495\textwidth]{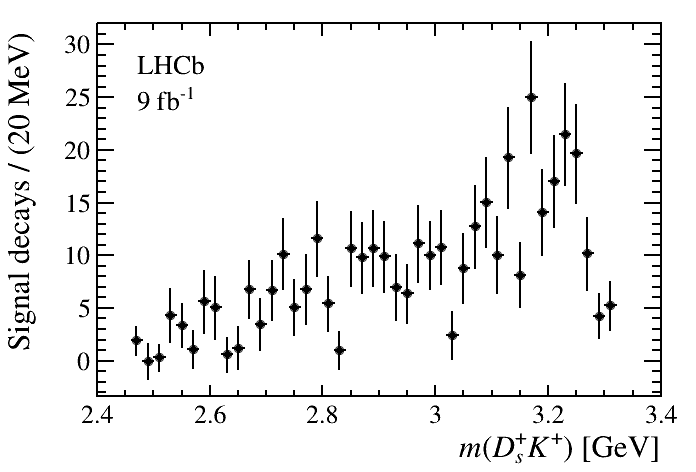}
    \caption{Distributions of background-subtracted invariant mass  of the (top)~\Dsp\Dsm, (bottom left)~\Dsm\Kp and (bottom right)~\Dsp\Kp systems of the \BuDsDsK decay.}
    \label{fig:proj_sig}
\end{figure}

To demonstrate the potential contributions from intermediate resonances in the \BuDsDsK decay, the invariant-mass distributions of the \Dsp\Dsm, \Dsm\Kp and \Dsp\Kp combinations, denoted respectively as $m(\Dsp\Dsm)$, $m(\Dsm\Kp)$ and $m(\Dsp\Kp)$, are shown in Fig.~\ref{fig:proj_sig}. The combinatorial background is subtracted using the \sPlot method~\cite{Pivk:2004ty} with the \Bp candidate mass exploited as the discriminating variable. A clear peaking structure is seen at the \Dsp\Dsm mass threshold, corresponding to a charmonium(-like) candidate. An amplitude analysis is performed to study the resonant contributions in the \BuDsDsK decay, presented separately in Ref.~\cite{LHCb-PAPER-2022-018}.
The intermediate resonant contributions to the \BuDDK decay have been studied in the previous LHCb analyses described in Refs.~\cite{LHCb-PAPER-2020-024,LHCb-PAPER-2020-025}.

Pseudo-experiments are carried out to check potential biases on the extracted signal yields in the signal and normalisation channels. Numerous samples are randomly generated according to the determined PDFs where the number of signal~(background) candidates in each sample is varied according to a Poissonian distribution with the mean set to the determined signal~(background) yield, referred to as the default result. The signal yield in each generated sample is determined by a fit to the \Bp candidate mass distribution and its variation from the default result is quantified as the pull value, $(N_i-N_0)/\sigma_i$, where $N_i$ and $\sigma_i$ are the signal yield and its statistical uncertainty in the $i$-th sample, respectively, and $N_0$ is the default signal yield. The pull distribution is then fitted using a Gaussian function. The mean represents the bias on the signal yield relative to the size of its statistical uncertainty. The bias is $\delta_{\rm sig}=0.09\pm0.01$ for the signal channel and is corrected for in the branching fraction calculation. Here, the uncertainty is due to limited number of the pseudo-experiments. No obvious bias is found on the signal yield for the normalisation channel.

The measured \BuDsDsK and \BuDDK signal yields contain residual NDC contributions, which need to be subtracted in the branching fraction calculation presented in Sec.~\ref{sec:br}.
The expected NDC yields are estimated using the same method as that in Refs.~\cite{LHCb-PAPER-2020-024,LHCb-PAPER-2020-025}, based on the number of candidates in the $D_{(s)}^\pm$ mass sidebands, which are 30--80\mev away from the known $D_{(s)}^\pm$ mass~\cite{PDG2020}. The fractions of the NDC yields in the total \Bu signal yields, are found to be \mbox{$f_{\rm NDC}^{\rm sig}=(5.2\pm2.7)\%$} and \mbox{$f_{\rm NDC}^{\rm norm} = (3.2\pm0.6)\%$} for the \BuDsDsK and \BuDDK channels, respectively. The uncertainties are due to the limited number of the candidates in the $D_{(s)}^\pm$ mass sidebands. The larger NDC-background contamination in the \BuDDK channel compared with that in Refs.~\cite{LHCb-PAPER-2020-024,LHCb-PAPER-2020-025} is due to relatively looser selection requirements in this paper.

\section{Branching fraction calculation}
\label{sec:br}
\begin{figure}[!tbp]
    \centering
    \includegraphics[width=0.495\textwidth]{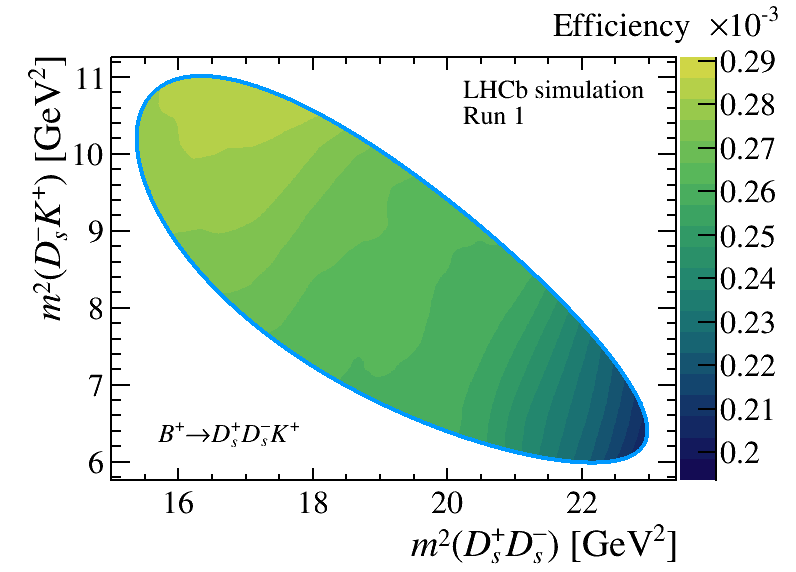}
    \includegraphics[width=0.495\textwidth]{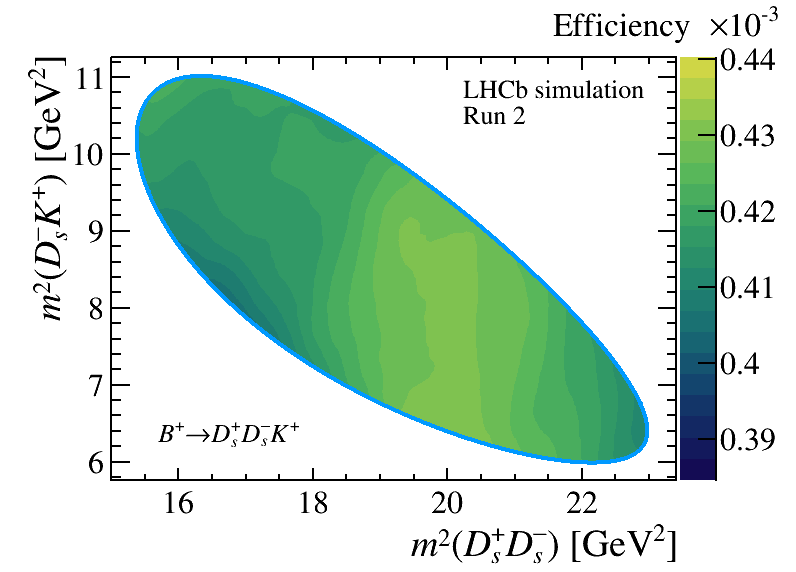}
    \includegraphics[width=0.495\textwidth]{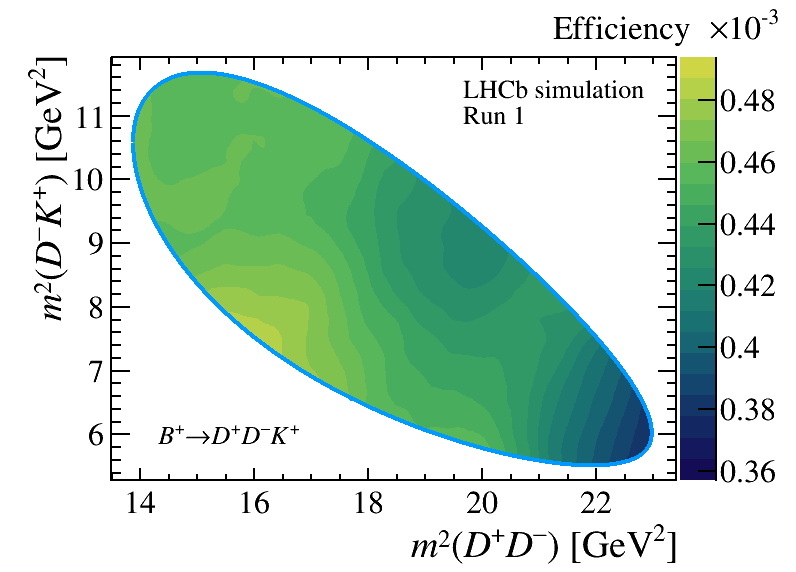}
    \includegraphics[width=0.495\textwidth]{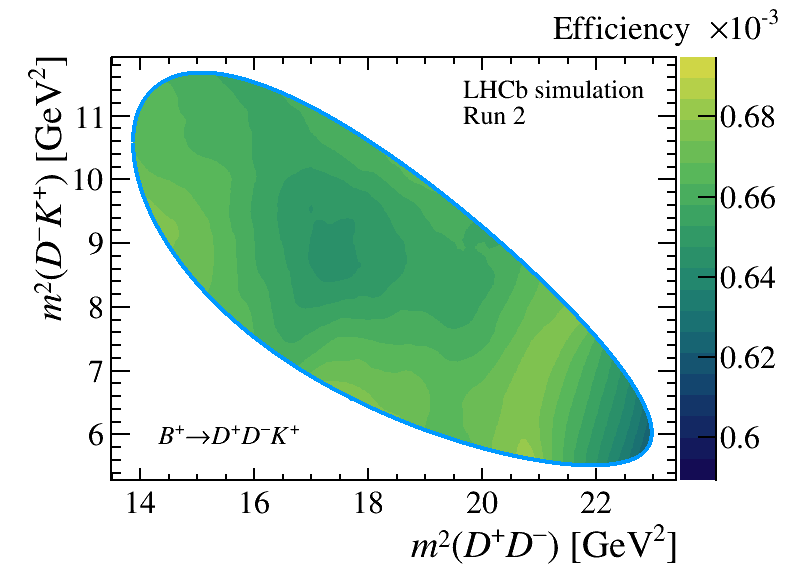}
    \caption{Efficiency as a function of the Dalitz-plot variables for the (top)~\BuDsDsK and (bottom)~\BuDDK decays, calculated separately for the (left)~Run1 and (right)~Run2 periods.}
    \label{fig:eff_kde}
\end{figure}

The branching fraction of the \BuDsDsK decay is measured relative to that of the \BuDDK decay,  
\begin{align}
    \Rbr \equiv \Rfrac &=  {\Nsigcorr\over\Nconcorr} \ \left[{\BrD \over \BrDs}\right]^2 \, ,
    \label{eq:R}
\end{align}
where the branching fractions of the decays \BuDDK, $\Dp\to\Km\pip\pip$ and $\Dsp\to\Km\Kp\pip$ are taken from Ref.~\cite{BaBar:2010tqo,PDG2020}. The efficiency-corrected signal yields of the \BuDsDsK and \BuDDK channels,  \Nsigcorr and \Nconcorr, are determined as 
\begin{align}
    \Nsigcorr &= \sum_{i} {\wsigi \over \esigi(\msDsDs, \msDsK)}\, , \\
    \Nconcorr &= \sum_{i} {\wconi \over \econi(\msDD, \msDK)}\, ,  
    \label{eq:Nsigcorr}
\end{align}
where the index $i$ runs over all the selected \Bp candidates in the signal or normalisation channel. 
The weight \wsigi or \wconi assigned to each candidate is obtained using the \sPlot method~\cite{Pivk:2004ty}; summing over these weights effects a statistical subtraction of the combinatorial background contribution. The discriminating variable is the reconstructed \Dsp\Dsm\Kp or \Dp\Dm\Kp mass, as shown in Fig.~\ref{fig:fit_sig}.
The efficiencies \esigi and \econi, determined using simulated samples, take into account the effects of geometric acceptance, reconstruction and selection requirements. They are evaluated separately in the two data-taking periods, 2011--2012~(Run1) and 2015--2018~(Run2), because of the differences in the collision energy and the detector configuration, and are shown in Fig.~\ref{fig:eff_kde} as functions of the Dalitz-plot variables of the corresponding decays. The kernel density estimation~(KDE)~\cite{Poluektov:2014rxa} technique is employed to obtain smooth efficiency distributions across the Dalitz plots. 
Finally, the efficiency-corrected yields are determined to be 
$\Nsigcorr=(9.5\pm0.6)\times 10^5$ and $\Nconcorr=(5.33\pm0.11)\times10^6$, where the uncertainties are statistical only.

Before obtaining the branching-fraction ratio $\Rbr$, it is necessary to subtract the NDC background contributions and to correct for the small fit bias on the extracted signal yield in the \BuDsDsK channel. 
These two effects are accounted for by introducing in Eq.~\ref{eq:R} two correction factors: $(1-f_{\rm NDC}^{\rm sig})/(1-f_{\rm NDC}^{\rm norm})$ for the NDC background subtraction, and $1-{\sigma_{\rm sig}\cdot \delta_{\rm sig}/ N_{\rm sig}}$ for the bias correction, where $N_{\rm sig}\pm \sigma_{\rm sig}=360\pm22$ and $\delta_{\rm sig}=0.09$. The branching-fraction ratio is determined to be 
\begin{equation}
    \Rbr = 0.525\pm0.033,
    \label{eq:br_stats}
\end{equation}
where the uncertainty is due to statistical uncertainties on the efficiency-corrected yields of both the \BuDsDsK and \BuDDK channels.

\section{Systematic uncertainties}
\label{Sec:sys}
\begin{table}[tbp]
    \centering
    \caption{Sources of systematic uncertainty on the branching-fraction ratio. The statistical uncertainty is also given.}\label{tab:sys_br}
    \begin{tabular}{lc}
        \hline
        Systematic source & Relative uncertainty~(\%) \\ \hline
        Signal model & 0.3 \\
        Background model & 0.1 \\
        NDC background & 2.9 \\
        Mass fit bias & 0.1 \\
        Multiple-candidate removal & 0.7 \\
        Hardware trigger correction & 2.3 \\
        PID correction & 2.8 \\
        Classifier modelling& 1.6 \\
        Tracking & 1.0 \\
        Truth matching & 0.6 \\
        Limited size of simulated samples & 0.5 \\
        KDE parameters & 0.4 \\
        \hline
        Total syst.~(stat.) & \phsys5.1~(6.3) \\
        \hline
    \end{tabular}
\end{table}

The sources of the systematic uncertainty on the branching-fraction ratio are summarised in Table~\ref{tab:sys_br}.
They can be classified into two categories: the effects on the signal yields and the effects on the efficiencies. The total uncertainty is obtained by adding the contributions in quadrature.

The sources affecting the signal yields include the choice of the signal and background PDFs in the \Bu mass fits, the uncertainties on the NDC background fractions in the signal and normalisation channels, the uncertainty on the bias of the \BuDsDsK signal yield and the multiple-candidate removal strategy. 
To evaluate the systematic uncertainty related to the choice of the signal PDF, an alternative model with a Hypatia function~\cite{hypatia} is employed in the \Bu mass fits to the signal and normalisation channels. The effect due to the choice of the background PDF is evaluated in a similar way, where a second-order polynomial function is used as an alternative. The NDC background fractions in the signal and normalisation channels have large uncertainties, which are due to the limited size of the data samples in the \Dpm and \Dspm mass sidebands. These uncertainties are propagated to the branching-fraction ratio. The uncertainty on the bias of the \BuDsDsK signal yield, resulting from the limited number of the pseudo-experiments that are employed to estimate the bias, is propagated in a similar way. Finally, the effect of the multiple-candidate removal strategy is evaluated by retaining an arbitrarily chosen candidate in an event.

The remaining systematic uncertainty sources in Table~\ref{tab:sys_br} influence the efficiencies in the signal and normalisation channels.
The hardware trigger is generally not well modelled in the LHCb simulation but the effect is expected to largely cancel in the branching-fraction ratio thanks to the similarity in the topology between the signal and normalisation channels. The residual effect is evaluated by applying to the simulated events an event-by-event correction that is determined by calibration samples~\cite{MartinSanchez:1407893}. 

The PID response is not well described in the simulation either and thus is corrected in the default result, with the correction determined from calibration samples using the PID transformation approach~\cite{LHCb-PUB-2016-021,LHCb-DP-2018-001}. The performance of the correction is limited by the size of the calibration samples, the choice of the values of hyperparameters in the transformation approach and the choice of the correction approach itself. 
The corresponding effects are investigated by extracting the efficiency maps from simulated samples with the correction determined by using alternative calibration samples bootstrapped from the original ones, varying the values of the hyper parameters of the PID transformation approach, and choosing the PID resampling approach~\cite{LHCb-PUB-2016-021,LHCb-DP-2018-001}. 
The resulting changes on the branching-fraction ratio are taken as the systematic uncertainties and are added in quadrature to obtain the uncertainty due to the imperfect PID correction.

The potentially imperfect descriptions of the discriminating variables in the simulation can cause bias on the efficiencies of the BDTG classifiers. The effect is evaluated by varying the BDTG selection for the two channels and the resulting change of the branching-fraction ratio is taken as the systematic uncertainty. 

The uncertainties on the tracking efficiencies for the same species of final-state particles in the signal and normalisation channels are expected to be the same and fully correlated, so their effects on the branching-fraction ratio largely cancel. The residual effect is from different final-state particles in the two channels, \ie\ $\Dsp\to\Km\Kp\pip$ and $\Dp\to\Km\pip\pip$. The systematic uncertainty on the branching-fraction ratio is evaluated as $1.0\%$~\cite{LHCb-PAPER-2018-036,LHCB-PAPER-2020-021}.

In the simulation, a truth-matching algorithm is used to identify signal decays. The efficiency of this algorithm is not $100\%$, leading to an underestimation of the efficiencies shown in Figure~\ref{fig:eff_kde}. However, the truth-matching efficiencies are expected to be very similar between the signal and normalisation channels, and thus the associated systematic uncertainties largely cancel. To evaluate the potential residual effect, the truth-matching efficiencies in the two channels are estimated using \Bu mass fits to the simulated samples with and without truth matching. The efficiencies are introduced as correction factors to the branching-fraction ratio, and the resulting variation is taken as the systematic uncertainty.  

The kernel width in the KDE method is chosen to be $2\gev^2$ in the default result. To evaluate the systematic effect due to this choice, two alternative values, $1\gev^2$ and $3\gev^2$, are taken in the efficiency parameterisations. The maximum change of the branching-fraction ratio is taken as the systematic uncertainty. 

The last systematic uncertainty source affecting the efficiencies is the limited size of simulated samples. The uncertainty is propagated to the branching-fraction ratio using the bootstrap method~\cite{efron:1979}. Bootstrapped samples are produced by randomly selecting from the simulated candidates with replacement. Each such sample has the same size as the original simulated sample.  These samples are employed to determine efficiencies and finally to obtain a set of branching-fraction ratios, whose standard deviation is taken as the systematic uncertainty.

\section{Summary}
\label{sec:summary}
In conclusion, the \BuDsDsK decay is observed for the first time using the proton--proton collision data collected by the LHCb detector at centre-of-mass energies of 7, 8 and 13\tev, corresponding to an integrated luminosity of $9\invfb$. The significance of the signal observed is larger than $10\,\sigma$.
The branching fraction of this decay is measured relative to that of the \BuDDK normalisation channel as 
\begin{equation}
    \Rbr ={\Brsig\over\Brcon} =  0.525 \pm 0.033 \pm 0.027 \pm 0.034.
    \label{eq:br_conclusion}
\end{equation}
With $\Brcon=(2.2\pm0.7)\times10^{-4}$ as the input~\cite{BaBar:2010tqo,PDG2020}, the \BuDsDsK branching fraction is then determined to be 
\begin{equation}
\Brsig = (1.15 \pm 0.07 \pm 0.06 \pm 0.38)\times 10^{-4}.
\end{equation}
In each of these results, the first and second uncertainties are statistical and systematic, respectively, and the third is from the uncertainties on the known branching fractions of the $\Dp\to\Km\pip\pip$~\cite{CLEO:2013rjc}, $\Dsp\to\Km\Kp\pip$~\cite{PDG2020,BaBar:2010ixw,CLEO:2013bae,Belle:2013isi} and \BuDDK decays~\cite{BaBar:2010tqo}.

The observation of the \BuDsDsK decay fills a major experimental gap in knowledge of the Cabibbo-favoured $\bquarkbar \to \cquarkbar \cquark \squarkbar$ transition with two open-charm hadrons in the final states~\cite{PDG2020}.
The measurement of its branching fraction is of great importance for investigations of the intermediate resonant states. In particular, it is vital input to studies probing the nature of the near-threshold structure in the \Dsp\Dsm system~\cite{LHCb-PAPER-2022-018}.


\section*{Acknowledgements}
%
%
\noindent We express our gratitude to our colleagues in the CERN
accelerator departments for the excellent performance of the LHC. We
thank the technical and administrative staff at the LHCb
institutes.
We acknowledge support from CERN and from the national agencies:
CAPES, CNPq, FAPERJ and FINEP (Brazil); 
MOST and NSFC (China); 
CNRS/IN2P3 (France); 
BMBF, DFG and MPG (Germany); 
INFN (Italy); 
NWO (Netherlands); 
MNiSW and NCN (Poland); 
MEN/IFA (Romania); 
MICINN (Spain); 
SNSF and SER (Switzerland); 
NASU (Ukraine); 
STFC (United Kingdom); 
DOE NP and NSF (USA).
We acknowledge the computing resources that are provided by CERN, IN2P3
(France), KIT and DESY (Germany), INFN (Italy), SURF (Netherlands),
PIC (Spain), GridPP (United Kingdom), 
CSCS (Switzerland), IFIN-HH (Romania), CBPF (Brazil),
Polish WLCG  (Poland) and NERSC (USA).
We are indebted to the communities behind the multiple open-source
software packages on which we depend.
Individual groups or members have received support from
ARC and ARDC (Australia);
Minciencias (Colombia);
AvH Foundation (Germany);
EPLANET, Marie Sk\l{}odowska-Curie Actions and ERC (European Union);
A*MIDEX, ANR, IPhU and Labex P2IO, and R\'{e}gion Auvergne-Rh\^{o}ne-Alpes (France);
Key Research Program of Frontier Sciences of CAS, CAS PIFI, CAS CCEPP, 
Fundamental Research Funds for the Central Universities, 
and Sci. \& Tech. Program of Guangzhou (China);
GVA, XuntaGal, GENCAT and Prog.~Atracci\'on Talento, CM (Spain);
SRC (Sweden);
the Leverhulme Trust, the Royal Society
 and UKRI (United Kingdom).




\addcontentsline{toc}{section}{References}
\bibliographystyle{LHCb}
\bibliography{main,standard,LHCb-PAPER,LHCb-DP,LHCb-CONF}

\newpage
\centerline
{\large\bf LHCb collaboration}
\begin
{flushleft}
\small
R.~Aaij$^{32}$\lhcborcid{0000-0003-0533-1952},
A.S.W.~Abdelmotteleb$^{50}$\lhcborcid{0000-0001-7905-0542},
C.~Abellan~Beteta$^{44}$,
F.~Abudin{\'e}n$^{50}$\lhcborcid{0000-0002-6737-3528},
T.~Ackernley$^{54}$\lhcborcid{0000-0002-5951-3498},
B.~Adeva$^{40}$\lhcborcid{0000-0001-9756-3712},
M.~Adinolfi$^{48}$\lhcborcid{0000-0002-1326-1264},
H.~Afsharnia$^{9}$,
C.~Agapopoulou$^{13}$\lhcborcid{0000-0002-2368-0147},
C.A.~Aidala$^{77}$\lhcborcid{0000-0001-9540-4988}
% ...

{\footnotesize \it

% ...
$^{9}$Universit{\'e} Clermont Auvergne, CNRS/IN2P3, LPC, Clermont-Ferrand, France\\
% ...
$^{13}$LPNHE, Sorbonne Universit{\'e}, Paris Diderot Sorbonne Paris Cit{\'e}, CNRS/IN2P3, Paris, France\\
% ...
$^{32}$Nikhef National Institute for Subatomic Physics, Amsterdam, Netherlands\\
% ...
$^{40}$Instituto Galego de F{\'\i}sica de Altas Enerx{\'\i}as (IGFAE), Universidade de Santiago de Compostela, Santiago de Compostela, Spain\\
% ...
$^{44}$Physik-Institut, Universit{\"a}t Z{\"u}rich, Z{\"u}rich, Switzerland\\
% ...
$^{48}$H.H. Wills Physics Laboratory, University of Bristol, Bristol, United Kingdom\\
% ...
$^{50}$Department of Physics, University of Warwick, Coventry, United Kingdom\\
% ...
$^{54}$Oliver Lodge Laboratory, University of Liverpool, Liverpool, United Kingdom\\
% ...
$^{62}$Syracuse University, Syracuse, NY, United States\\
% ...
$^{77}$University of Michigan, Ann Arbor, MI, United States, associated to $^{62}$\\
\bigskip
% ...
}
\end{flushleft}

\end{document}